# OPERATIONAL EXPERIENCE OF ALBA'S DIGITAL LLRF AT SOLARIS LIGHT SOURCE

P. Borowiec[†], L. Dudek, W. Kitka, A. Kisiel, P. Klimczyk, M. Knafel, M. Kopec, A. Wawrzyniak,
National Synchrotron Radiation Centre SOLARIS, Krakow, Poland,
F. Perez, A. Salom, ALBA-CELLS Synchrotron, Cerdanyola del Valles, Spain,
A. Andersson, R. Lindvall, L. Malmgren, A. Mitrovic, MAX IV Laboratory, Lund, Sweden

*Abstract*

For control of RF cavities installed in Solaris storage ring light source the digital Low Level RF (dLLRF) system was necessary from the beginning of operation. Since there were no expertise at the new constructed facility and no time for development due to funds deadline, almost turn-key dLLRF from Alba has been implemented according to MAXIV selection. Thanks to high flexibility of dLLRF only small adaptations were needed in terms of interfaces to auxiliary systems and setup of parameters. This paper summarizes operational experience about installation, commissioning, learning-curve from entry-level user, beam operation and future upgrades of this dLLRF.

## INTRODUCTION

Solaris in Krakow – the first Polish synchrotron light source was started an operation with the beam in storage ring from May 2015. The storage ring (SR) and related components are a twin brother of MAXIV 1.5 GeV SR installed in Lund, Sweden. The facility consists of a 1.5 GeV, 96 m circumference SR and a 600 MeV S-band linear accelerator with RF thermionic gun. Since linac is not a full energy one, ramping in SR is necessary. The RF systems of the SR work at 99.93 MHz. There are two normal conducting, capacity loaded accelerating cavities and two 3rd order Landau passive cavities. Active cavities are fed by 60 kW SSA each with 120 kW isolator in series. All cavities are controlled from one dLLRF system.

## DIGITAL LOW LEVEL RF HARDWARE

Main components of the dLLRF system are:

Digital hardware: commercial uTCA chassis with MCH from Vadatech; 2 pieces of commercial uTCA boards with a Virtex-6 FPGA mother board (Perseus 601X), 1st with double stacked FMC boards with fast ADCs (MI125 with 16 ADCs – 14 bits – 125 MHz) and DACs (MO1000 with 8 DACs – 14 bits – 1000 MSPS) called Loops Board which processes RF signals needed for the main feedback loop; and 2nd with fast ADCs (MI125 with 16 ADCs – 14 bits – 125 MHz) called Diagnostic Board which processes RF signals for interlock purposes. Additionally 2 Mestor breakout boxes provide 32 GPIO bus and 4 slow ADCs each. All uTCA boards came from Nutaq. Linux based host computer in the uTCA chassis runs Device Servers for Tango control system and it allows for local access to the system. FPGA boards interchange information from/to the Host PC through the uTCA backplane. Whole system is connected through Ethernet to the facility network. Separate Digital Patch Panel is the interface between high density connectors from Mestor to other signals from the machine and match signals levels and impedances. The capabilities of FPGAs allowed including the control of all cavities, the handling of fast interlocks, providing automatic start up and conditioning of cavities, and post-mortem analysis in one dLLRF system.

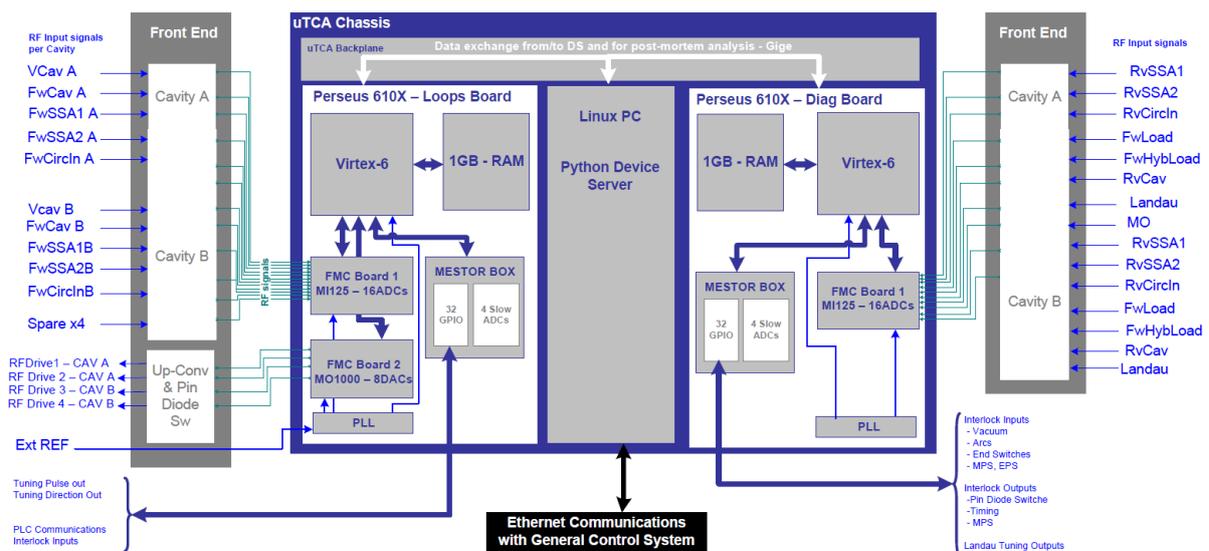

**Figure 1:** Hardware diagram of dLLRF[1]

† email address: pawel.borowiec@uj.edu.pl

Analog Front Ends: Loops Front End for RF drive up-conversion and signals conditioning. It consists also: VCXO with PLL for LO generation and FPGA clock synchronization with MO reference; and PIN diode switches for output RF control. Diagnostic Front End is for signals conditioning. Both front ends provide BNC test points of all RF signals.

Detailed hardware description is presented in [1].

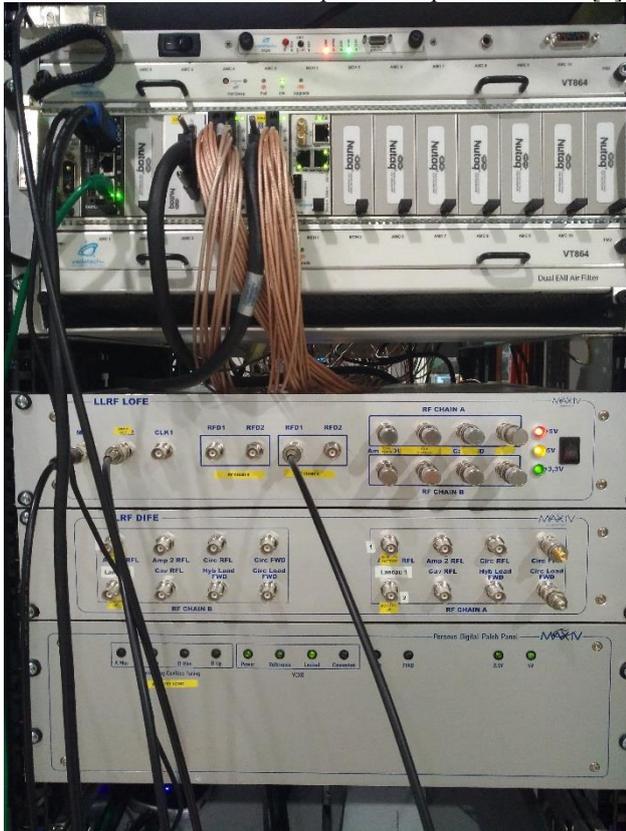

**Figure 2:** Hardware implementation of the dLLRF in SOLARIS

# FUNCTIONALITIES AND OPERATIONAL EXPERIENCE

## Feedback loop for active cavities

It consists PI algorithm implemented in FPGA, based on I, Q demodulation technique for voltage and phase control of the cavity. In the loops following RF signals are used: cavity voltage, forward RF power from different points of the system, Master Oscillator. Input for the loop can be selected, depends on the needs. That possibility was used during conditioning of active cavities when forward power to the cavity was stabilized (directional coupler port) instead of the field in the cavity (pick-up probe) to avoid huge RF transmitter output power fluctuations in case of field break down due to arcing.

First, system has been setup for operation without beam. The dLLRF parameters allow for attenuation (open loop gain) and phase (phase shift of ADC and DAC) compensation of RF cables. For that reason RF cables from patch panel to the cavities are not precisely phase matched. Adjustment was performed during commission and it is necessary after restart of the system or hardware changes. Then parameters of the RF power increase rate and PI controller have been set. Kp parameter is kept at minimal value and Ki at half of the value when system becomes unstable. Operation with the beam was started with only one cavity controlled, second one was detuned. It simplified commissioning stage of storage ring when it wasn't needs of more RF power. Afterwards phase matching of the 2 active cavities was necessary. It was done at the first stage by observation of forward power from RF transmitters, after calibration of the system by calculation of the synchronous phase, which is 167.4°±2.7 [2] Long term test shows feedback loop allows to stabilize the phase in cavity within ±0.5 peak to peak (0.17°RMS) and amplitude within ±2 % peak to peak (0.2 RMS).

## Tuning loop for active cavities

It allows to keep resonance frequency of the cavity by comparison of the phase of forward power to the cavity and the cavity voltage. When the phase difference is outside of the definable region (called tuning dead-band) cavity is tuned by elastic deformation of its end-plate. The dead-band prevents oscillation of the stepper motor and additional delay between successive tuner movements extends tuning mechanism lifetime. The working point of the cavity is set 10° away from the optimum point to have enough room from negative slope of the resonance where tuning may become unstable. Besides tuning loop the manual tuning of all cavities is available. This is usable for tuning/detuning of the cavity when needed.

## Automatic Start-up

This utility allows the dLLRF to recover smoothly after an interlock. When the dLLRF detects any interlock or the transmitter is not ready, the Loops board of the dLLRF will go to Standby state. In this case, the IQ Loops are disabled and the RF Drive of the dLLRF is equal to the definable minimum level. Also the tuning loop remains in standby state since the measured forward power of the cavity is below the definable threshold. Once the Tx is enabled and dLLRF has detected minimum RF Power, it tunes the cavity and just after that, feedback loop is enabled. Then dLLRF increases smoothly the RF Drive from low to demanded set point with defined increase rate. When set point is achieved, the dLLRF is ready to start operation with the beam. Since delivery of the RF Power to the cavity in Solaris is controlled by enabling of the RF transmitter that feature allows for operators to manage complex procedure with just one button.

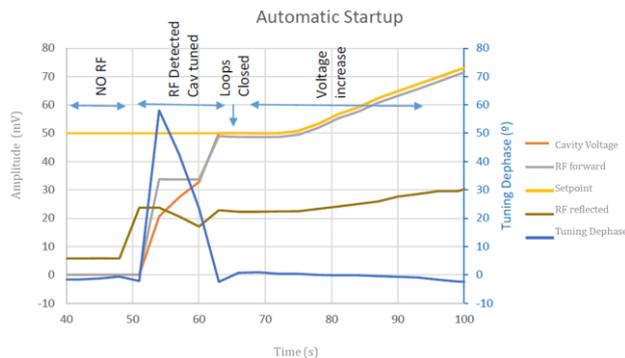

**Figure 3:** Automatic start-up process[3]

*Automatic Conditioning*

That feature is used to speed up the conditioning process of the cavities. It was used especially during first power up of the cavities after installation in storage ring and after works on cavities when the vacuum was break down. The three digital outputs of a Vacuum Gage Controller are connected to the GPIO bus of the dLLRF. These outputs are dry contacts that get opened when the pressure of the cavity is above certain limits. At Solaris low level is set to 1e-8mbar and high level to 5e-8mbar. When the pressure is below both levels, the automatic conditioning allows increasing the RF power. If the vacuum level is above higher pressure limit, the RF drive remains constant until the pressure is again below the lower threshold level. Third output is used as hard interlock to stop RF Power delivery at pressure above 1e-7 mbar. Additionally the RF Drive can also be square modulated at 10 Hz. The operator can adjust the duty cycle of the modulation. Pulse conditioning helps for vacuum recovery because ion sputter pump has more time for pump down. That capability was used in the multipacting region.

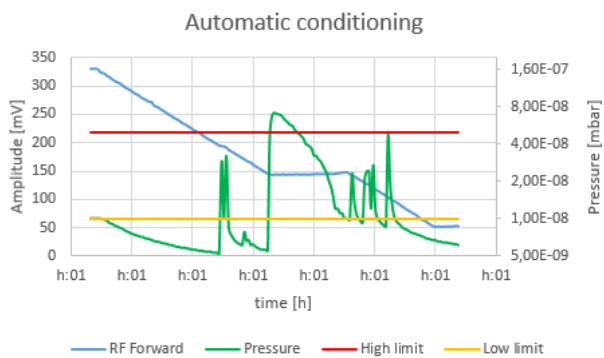

**Figure 4:** Automatic conditioning during RF power ramp down

Usually automatic conditioning is used during ramp up of the RF power but on the [Fig. 3] is presented its action during power ramp down. Cycling of the RF power and phase in whole range was used for detection of unconditioned areas in the cavity. They appeared usually around multipacting region which could be crossed during ramp up of the RF power but it could appear again during cycling. As presented, dLLRF has limited change of RF forward power when pressure crossed so called high limit and decreased RF power when pressure was below low limit according to set point given by cycling software.

*Field stabilization in Landau cavities*

Although Landau cavities are passive their field can be stabilized by tuning with elastic deformation of end plates. When cavity voltage is out of defined range expressed in percent of the set point the cavity is tuned. Maintaining of the constant voltage ratio of active cavities to Landau is a key for the beam lifetime. Since Solaris storage ring is operating in decay mode it turned out during test of the algorithm that cavity is tuned close to resonance where instabilities blow up the beam. Recently new parameter has been added, stabilization is active only above minimum beam current. It requires beam current information directly in dLLRF hardware, the interface is under production.

*Fast Interlocks*

Many signals from different points of the RF system are monitored (cavity, isolator input and output, amplifiers output) and other signals from equipment like Machine Protection System, Vacuum interlocks, Limit Switches of cavity motors, Transmitter interlocks and Arc Detectors. Whenever the signal at any of these points reaches a defined threshold or any other interlock is detected, the Diagnostics Board is opening in microseconds the PIN diode switches which are in series with RF Drive. In parallel Loops Board is set in standby mode (loops disable and DACs input set to minimum values) and the PLC controlling the RF amplifier is informed about the state. Interlocks are stored in order of appearance for further analysis.

Logic interlocks of different types (dry contact, TTL, 50 Ω) from many locations are collected in Digital Patch Panel. During commissioning false interlocks from arc detectors (TTL signal) in the circulator have been triggered during powering of RF transmitters. Although inputs of Digital Patch Panel are opto-isolated but their cathodes are connected to the common ground where problem was identified. Additional opto-couplers in arc detector output solved the issue.

*Fast Data Logger (FDL)*

. After presence of the interlock the FDL can be triggered. This utility consists of a 1GB RAM (around 0.5 seconds of operation) continuously acquiring data in a circular buffer. When a trigger is raised either on demand or due to an interlock, the acquisition is stopped and the RAM data is sent through the uTCA backplane to the host PC for post-mortem analysis.

Currently FDL is only implemented to download data after interlock, additional High Level Software programming is needed for data acquisition before interlock, which is the most interesting.

*External communication*

Tango control system is used in Solaris. Device Servers are running on the host computer for loop and diagnostics boards which allow exchange of data with FPGAs. Operators are using GUI in control room for everyday operation.

Even after crash of the communication, device servers or host computer, beam is kept because FPGA boards are still running. Only disadvantage is that there is no data flow to control system. Host computer allows for local operation with dLLRF and together with tools delivered from Nutaq (command line interface access to FPGA boards) it is a convenient package during debugging process.

## ONGOING WORKS

### Spare parts stock

Since one electronics controls whole RF system (4 cavities) there is significant risk to be out of operation in case of its malfunction. For that reason backup system: uTCA chassis, both Perseus boards and PLL with VCXO were ordered.

### In-house developments

Having complete spare system we can start with development of the new functionalities with possibility of tests independently from beam operation. Such examples are: beam current interface, limitation of RF transmitter forward power below maximum dissipated power of active cavities, separate amplification block for each feedback input. Recently training about dLLRF was given by A. Salom what is a next step in building of knowledge at Solaris.

## CONCLUSION

High functionality of the dLLRF gave quick start-up of RF system for storage ring. It covers not only RF control but provides extensive protection and diagnosis tools as well. System was delivered just in time when all machine cabling were already prepared according to received documentation. Most demanding at the beginning was right setup of all dLLRF parameters. There is no any hardware malfunction after 2.5 years of operation. Firmware of FPGA's can be modified without restriction what gives possibility for the new developments.

## ACKNOWLEDGEMENT

For Angela Salom from ALBA for her extensive support during start-up of the dLLRF and then for accurate solutions for popped-up issues. For Max IV colleagues: Lars Malmgren, Aleksandar Mitrovic, Robert Lindvall for their patience and sharing experience with cabling, operational and software stuff. For Solaris team members for signals delivery and integration in Tango control system.